\documentclass[%
reprint,
superscriptaddress,
 amsmath,amssymb,
 aps,
 pra,
]{revtex4-1}

\usepackage{graphicx}
\usepackage{dcolumn}
\usepackage{bm}
\usepackage{hyperref}
\usepackage[mathlines]{lineno}

\usepackage{siunitx}

\begin{document}
\bibliographystyle{apsrev}

\title{High efficiency cyclotron trap assisted positron moderator}
\author{L. Gerchow}
\affiliation{ETH Zurich, Institute for Particle Physics, Otto-Stern Weg 5,  CH-8093 Zurich, Switzerland}
\author{D.~Cooke}
\affiliation{ETH Zurich, Institute for Particle Physics, Otto-Stern Weg 5,  CH-8093 Zurich, Switzerland}
\author{S.~Braccini}
\affiliation{University of Bern, Albert Einstein Center for Fundamental Physics (AEC), Laboratory for High Energy Physics (LHEP), University of Bern, Sidlerstrasse 5, CH-3012 Bern, Switzerland.
}
\author{M.~D\"obeli}
\affiliation{ETH Zurich, Institute for Particle Physics,  Otto-Stern Weg 5,  CH-8093 Zurich, Switzerland}
\author{K.~Kirch}
\affiliation{ETH Zurich, Institute for Particle Physics,  Otto-Stern Weg 5,  CH-8093 Zurich, Switzerland}
\affiliation{Paul Scherrer Institute, CH-5232 Villigen, Switzerland}
\author{U. K\"oster}
\affiliation{Institut Laue Langevin, 6 rue Jules Horowitz, F-38042 Grenoble Cedex 9, France}
\author{A.~M\"uller}
\affiliation{ETH Zurich, Institute for Particle Physics,  Otto-Stern Weg 5,  CH-8093 Zurich, Switzerland}
\author{N. P. ~Van Der Meulen}
\affiliation{Paul Scherrer Institute, CH-5232 Villigen, Switzerland}
\author{C.~Vermeulen}
\affiliation{Paul Scherrer Institute, CH-5232 Villigen, Switzerland}
\author{A.~Rubbia}
\affiliation{ETH Zurich, Institute for Particle Physics,  Otto-Stern Weg 5,  CH-8093 Zurich, Switzerland}
\author{P.~Crivelli}
\email{crivelli@phys.ethz.ch}
\affiliation{ETH Zurich, Institute for Particle Physics,  Otto-Stern Weg 5,  CH-8093 Zurich, Switzerland}

\date{\today}
\begin{abstract}

We report the realisation of a cyclotron trap assisted positron tungsten moderator for the conversion of positrons with a broad keV- few MeV energy spectrum to a mono-energetic eV beam with an efficiency of \SI{1.8(2)}{\percent} defined as the ratio of the slow positrons divided by the $\beta^+$ activity of the radioactive source. This is an improvement of almost two orders of magnitude
compared to the state of the art of tungsten moderators.
The simulation validated with this measurement suggests that using an optimised setup even higher efficiencies are achievable.

\end{abstract}

\pacs{Valid PACS appear here}
\maketitle


\section{\label{sec:level1}Introduction}

The positron and its bound state with an electron, positronium, have found many applications in physics and chemistry \cite{CharltonBook, JeanBook}. In medicine, they are used for what is probably the best known application of anti-matter in ``everyday" life, the method of positron emission tomography (PET) \cite{townsend2004}. Relying on their unique sensitivity to the electronic environment, positrons serve in applied science for the characterisation of materials. For example they provide one of the most sensitive techniques to detect defect concentrations \cite{tuomisto2013}, they can be used to perform measurements of Fermi surfaces \cite{eijt2006,weber2015}, determine pore sizes in polymers \cite{jean2013} and nanoporous films \cite{gidley2006}. The number of applications is continuously increasing e.g. prompted by the growing complexity of advanced functional materials with multi-level porosity such as hierarchical zeolites and metal-organic frameworks  \cite{oPsNC,oPsMOFs}. Studies of positron and positronium interactions with matter are also a  vibrant field of research \cite{surko2005, laricchia2015}.

As systems made of two elementary particles with no sub-structure, positronium (Ps), including Ps-ions and Ps$_2$ molecules, are ideal for testing bound state QED \cite{karshenboim2005, fee1993, asai2014, PsIon, PsIon2, cassidyPs2}, fundamental symmetries \cite{vetter, yamazaki2010} and to search for new physics \cite{PRD2007}.
 Work is in progress to improve on the current results, to measure the effect of gravity on antimatter using Rydberg Ps \cite{cassidyPRL2016} and to form a Ps Bose-Einstein condensate \cite{Avetissian2014}.
 Positrons and positronium are also essential ingredients for the production of anti-hydrogen currently being studied at CERN \cite{ALPHA, ATRAP, ASACUSA, AEGIS, GBAR}.

The development of slow positron beams in the seventies greatly expanded the possibilities of this field \cite{PosBeamAppl}. More recently, an additional boost was given by the advent of buffer gas traps allowing for manipulation and storage of large positron plasmas \cite{surko2015}.

There are different ways to produce positrons which include the use of accelerators
\cite{Krause2008,Oshima2009, Perez2009}, nuclear reactors \cite{Falub2002,Hawari2011, Hugenschmidt2014} or ultra-intense short pulsed lasers \cite{Chen2009}.
The most common and compact solution is to use radioactive isotopes that are $\beta^+$ emitters such as $^{22}$Na \cite{surko2015}.
To form a slow positron beam the positrons from the broad keV to few MeV energy spectrum of the source have to be converted to a mono-energetic eV beam using moderators. Those can be divided into two classes: metals with a negative work function \cite{LynnAPL1985} and materials with very long diffusion lengths for positrons \cite{MillsAPL1986}.
The best work function based moderators are thin single crystalline tungsten foils or tungsten meshes with efficiencies of the order of \num{e-4} \cite{Saito2002}.
 The most efficient moderators rely on the long diffusion length of positrons in frozen rare gases, e.g. neon has a typical efficiency of $\epsilon = \num{7e-3}$ \cite{surko2015}.

In this paper, we present a scheme based on cyclotron trap assisted moderation that improves the amount of positrons available for the moderation process resulting in a  higher efficiency. 

\section{Principle of the cyclotron trap assisted moderation}

\begin{figure}[h!]
	\centering
	\includegraphics[width=.45\textwidth]{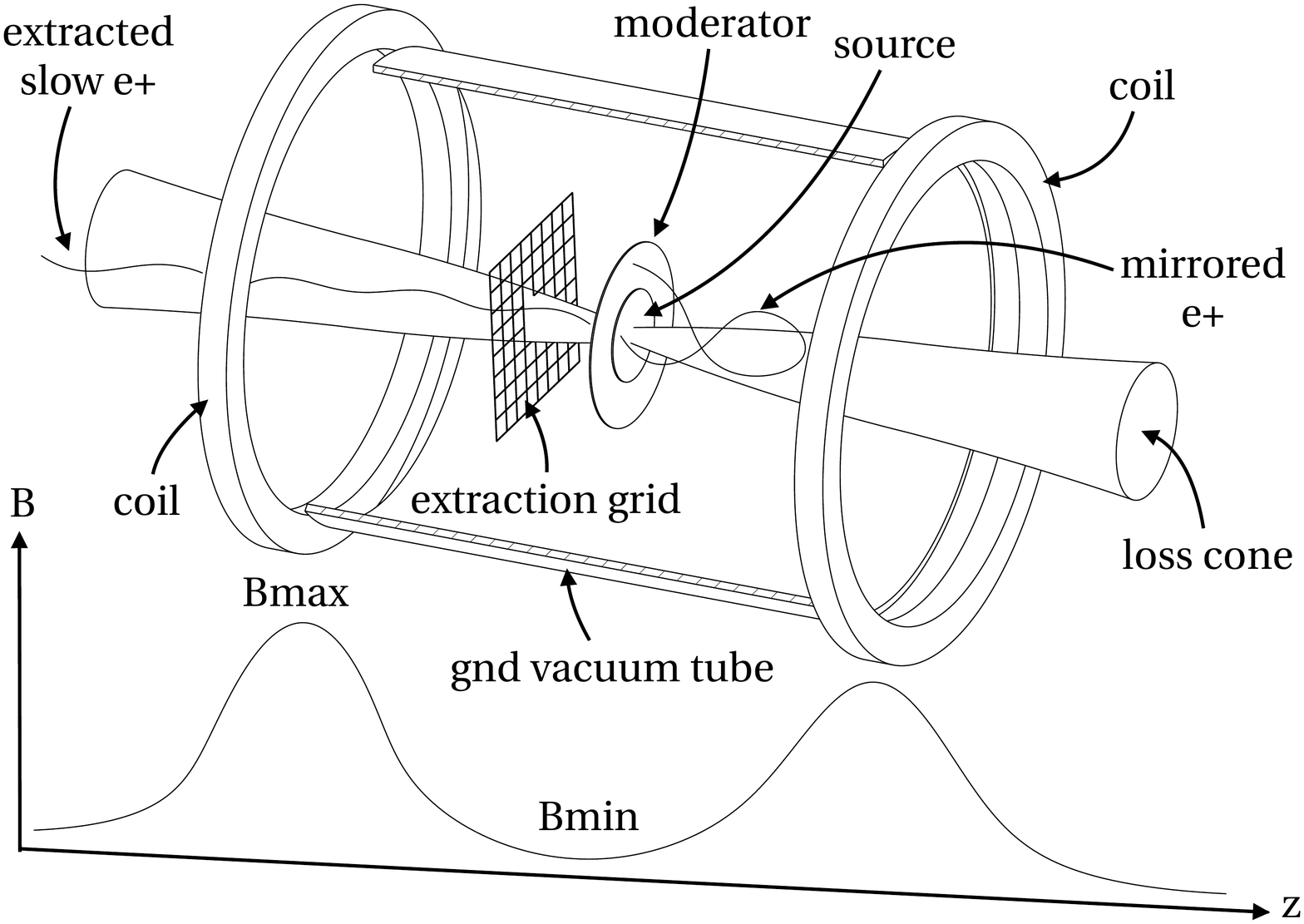}
	\caption{Scheme of the cyclotron assisted moderator principle. Two thin Ti activated foils ($^{48}$V), indicated as "source", and the W(110) foil, indicated as "moderator", placed inside a cyclotron trap act as a positron emitter, energy degrader and moderator. The confined positrons emitted from the source (kept at +100 V) lose energy passing through the foils until they are moderated. The use of a grid at ground potential maximizes the efficiency of the extraction of the moderated positrons.}
	\label{fig:HECTAM_scheme}
\end{figure}

A cyclotron trap (CT) is a magnetic bottle consisting of two coaxially identical coils separated by a given distance (see Fig. \ref{fig:HECTAM_scheme}). By running a current in the same direction through the coils, the created magnetic field along their central axis has a maximum value $B_\text{max}$ at the center of each coil and a local minimum $B_\text{min}$ between them. This leads to the confinement of charged particles if their momenta perpendicular $p_\perp$ and parallel $p_\parallel$ to the coil axis satisfy the relation (assuming adiabatic invariance):
\begin{equation}\label{eq:trapcondition}
	\left|{\frac{p_{\parallel 0}}{p_{\perp 0}}}\right| \leq \sqrt{\frac{B_\text{max}}{B_\text{min}}-1} \,.
\end{equation}
These particles then travel back and forth between the two coils on spiral trajectories along the trap axis.

In 1960, Gibson et al. \cite{Gibson1960}, reported the confinement of positrons in a ``mirror machine".
In the eighties \cite{Simons:1988fz,Eades:1989vk}, L. M. Simons proposed a CT with a thin foil placed in its middle to be used as an energy degrader for slowing down anti-protons and negative muons \cite{Simons:1993aa} to keV energies. This scheme has been used in the recent measurement of the proton charge radius with muonic-hydrogen \cite{Pohl2010}.
W. B. Waeber et al. \cite {Shi1994} tried to implement the same approach for degrading positrons to energies of a few keV before their extraction and subsequent moderation outside the CT to form a mono-energetic beam. However, due to their very challenging extraction scheme they could not reach the very promising theoretical predictions and this project was discontinued \cite{Gerola1995,Gerola1995NIMA}.

In the setup presented here, all the involved steps, i.e. the positron emission, the energy degradation, the moderation and the extraction are performed inside the cyclotron trap.  The source, an activated \SI{1}{\um} thick titanium foil, and the moderator, a \SI{1}{\um} thick single crystal tungsten (110) foil are matched and placed in the center of the cyclotron trap. This allows to greatly increase the amount of positrons available for moderation. The trapped positrons lose energy each time they pass through the foils. Once they have been degraded to energies of a few keV they thermalise in the foils and can be re-emitted as slow positrons. The very narrow energy spread, due to the negative work function of tungsten, guarantees that slow positrons can be extracted with an efficiency close to \SI{100}{\percent} from the trap (see Eq. \ref{eq:trapcondition}) when applying a small electric field.

\section{Experimental setup}
\begin{figure}[htb]
        \centering
        \includegraphics[width=.45\textwidth]{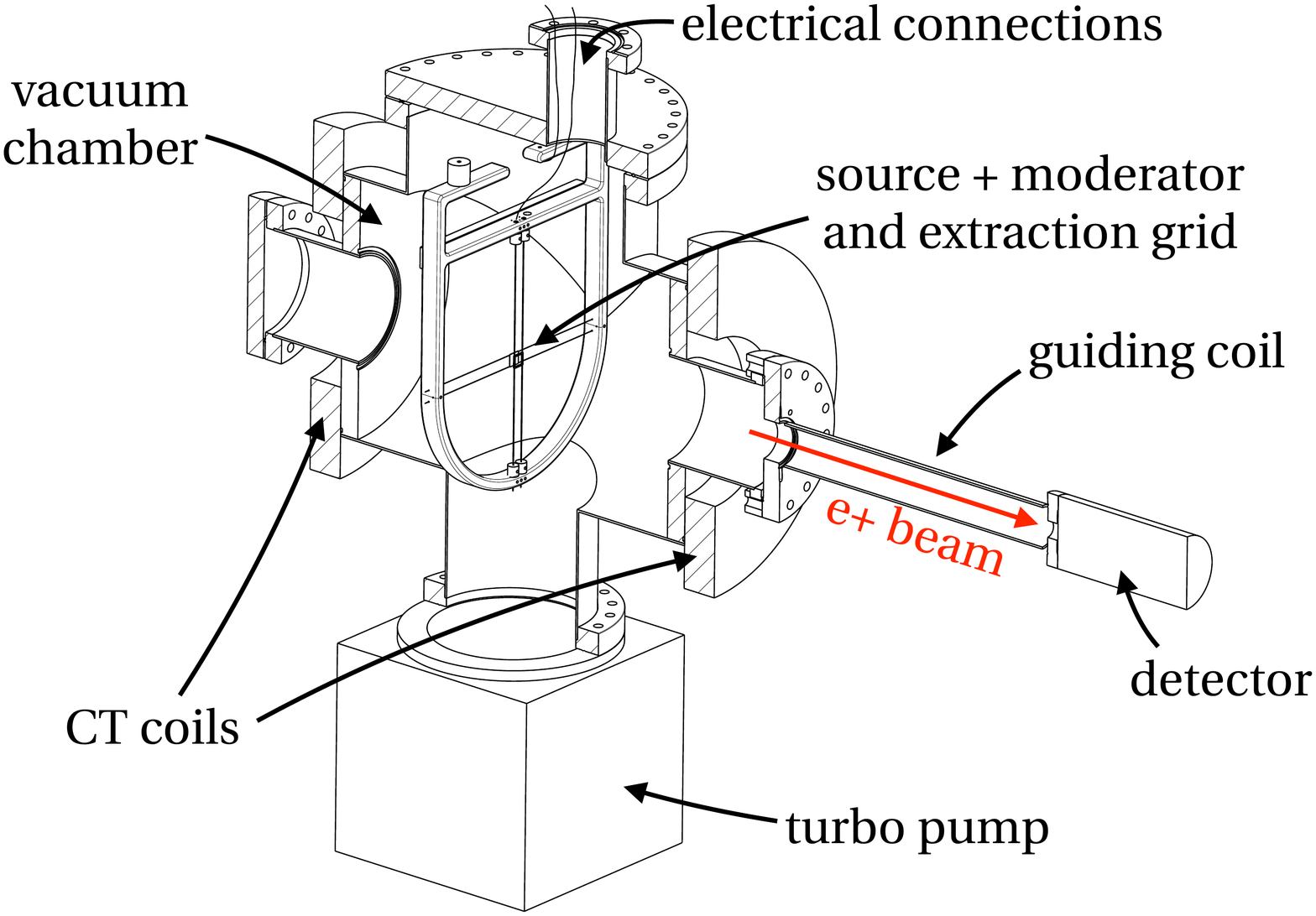}
        \caption{Schematic of the setup consisting of the vacuum chamber, the two coils forming the magnetic bottle, the guiding coil, the source, the moderator and the extraction grid mount, and the detector.}
        \label{fig:final_setup}
\end{figure}

The experimental setup is schematically illustrated in Fig. \ref{fig:final_setup}.
The sources for this experiment were $^{48}$V created by irradiation of \SI{1}{\um} Ti foils  exploiting the reaction $^{48}$Ti(p,n)$^{48}$V with 8 MeV protons from the ETHZ TANDEM accelerator and with 18 MeV protons using the external beam line of the IBA 18/9 medical cyclotron located at the Inselspital in Bern, Switzerland \cite{CAARIpaper}.

Calculations of the yield, confirmed by multiple Ti foil irradiations, show a good agreement for the end-of-bombardment activity $A_\text{EOB}$ in the thin target approximation:
\begin{equation}
	A_\text{EOB}= \text{ln}(2) \cdot \frac{\rho_\text{target} \cdot d \cdot \sigma(E)}{{t_{1/2}} \cdot m_{a}}  \cdot \Phi \cdot t
\end{equation}
with the target density $\rho_\text{target}$, thickness $d$, cross section $\sigma(E)$, projectile flux $\Phi$, irradiation time $t$, half-life $t_{1/2}$ and atomic mass $m_\text{a}$ of the target material. Those values are reported in Table \ref{sources_2}.

The half-life of $^{48}$V is 16 days which is not ideal for continuous operation of a positron beam. However, developments are underway that should allow one to achieve 10 MBq activities of the foils in less than 1 day irradiation using solid target stations usually employed for the production of radioisotopes with medical cyclotrons. Therefore, $^{48}$V is interesting for targeted measurements. For long term operation, better choices for the $\beta^+$ emitter are $^{58}$Co or $^{22}$Na with half-lives of 70 days and 2.6 years respectively.
 Preliminary attempt to produce $^{58}$Co via irradiation of $^{58}$Ni foils were performed at the SINQ spallation source of the Paul Scherrer Institute (PSI) but only a very limited activity of few kBq was achieved. In fact this reaction is suppressed for thermal neutrons and ideally one would use a fast reactor since the cross section for $^{58}$Ni(n,p) becomes appreciable only for neutrons above 0.5 MeV. Moreover,  $^{58}$Co has a high capture cross-section for thermal neutrons, thus in this case the produced positron emitting isotope is depleted.
The production of a $^{22}$Na source has a higher threshold (see Table \ref{sources_2}) and to achieve comparable activities requires about 1000 times more protons on target than the $^{48}$V production. The production of up to 20 MBq  $^{22}$Na sources by irradiation of  \SI{125}{\um} aluminum foils with 72 MeV protons has been demonstrated in the past at PSI \cite{beamNIMA,cugPhD}. With commercial 70 MeV cyclotrons for isotope production \cite{cyclotron2010}, 1 MBq sources on \SI{1}{\um} foils could be produced in 10 days of irradiation with  \SI{80}{\uA} currents. An important feature of activated thin metal foils is that those are vacuum compatible and the radioactive isotopes are well bound inside the foil and thus are ideal in terms of radiation safety.

\begin{table}[h]
\caption{\label{sources_2}{Endpoint energy $T_\text{max}$, half-life $t_{1/2}$, $\beta^+$ branching $\Gamma(\beta^+)$, production target and reaction (projectile,X), maximum cross-section $\sigma_\text{max}(E_\sigma)$, corresponding projectile energy $E_\sigma$, EOB activity $A_\text{EOB}$ and $\Phi\cdot t$ as \si{\uA}h/$e$ to create 1 MBq ($e$ C is the elementary charge) for \SI{1}{\um} foils of $^{58}$Co, $^{22}$Na and $^{48}$V \cite{NDS_IAEA}.}}
\begin{center}
\begin{tabular}{llll}
 Isotope &\textbf{$^{58}$Co} & \textbf{$^{22}$Na }& \textbf{$^{48}$V}\\
\hline
$T_\text{max} $ / keV &475  & 545  & 695 \\
$t_{1/2}$  & 70.85 d & 2.602 y & 15.97 d \\
$\Gamma(\beta^+)$  & 14.9\% & 90.6\% & 50\% \\
target(projectile,X)  & $^{58}$Ni(n$_\text{fast}$,p) & $^{27}$Al(p,X) & $^{48}$Ti(p,n) \\
$\sigma_\text{max}(E_\sigma)$  & 1-600 mb & 44 mb & 382 mb \\
$E_\sigma$ & 25 meV & 44 MeV & 12 MeV \\
$\rho_\text{target}$ / $\text{g}/\text{cm}^3$  & 8.9 & 2.7 & 4.5 \\
$A_\text{EOB}$ / kBq/(\si{\uA}h/$e$)  & 107 & 0.05 & 24 \\
$\Phi\cdot t$ / \si{\uA}h/$e$ for 1 MBq & 9.4 & 20000 & 41
\end{tabular}
\end{center}
\end{table}

The \SI{1}{\um} tungsten moderator was purchased from the Dept. of Physics and Astronomy of the University of Aarhus, Denmark and annealed $2\times 15$ minutes through electron bombardment shortly before mounting it in the setup. For the slow positron extraction a 96\% transmission tungsten mesh was used. The cyclotron trap is formed by two identical water cooled coils that were re-used from an experiment at PSI \cite{WEBmagnet} after removal of the iron yoke and refurbishing. 
 The maximal characteristic values of the CT running with a current of 650 A are $B_\text{max}=2.559(2)$ kG and $B_\text{min}=0.544(2)$ kG corresponding to a magnetic field ratio of 4.704(17).
The typical energy distribution of the moderator used in this experiment is  $\Delta E_{\parallel,\text{mod}}\approx1$
eV \cite{Vehanen:1983aa,firstmoderation,Appl.Phys.Lett.51.1862}. Due to this energy spread not all slow positrons can escape from the trap. In order to maximize the extraction efficiency a grid has been added to produce a 200 V/cm electric field between moderator and grid.

To reduce the background from the annihilation in the source and from unmoderated positrons that could reach the detector, a 1 m long 100 G solenoid guides the positrons away from the CT (see Fig. \ref{fig:final_setup}).

Due to the limited activity of the source available for this experiment we selected an electron multiplier (EM) to detect the slow positrons since it has smaller dark counts ($<$ 1 count/s) compared to micro-channel plate detectors (MCPs). The drawback is that EMs are very sensitive to magnetic fields and therefore the positrons have to be extracted from the guiding field used for their transport.
This was realized by terminating the magnetic field with Mu-metal shielding. Simulations with SIMION and COMSOL were used to optimise the extraction efficiency to values close to 100\%. This was confirmed experimentally by using the ETHZ slow positron beam \cite{beamNIMA}.

\section{Simulation} 
\label{sec:simulation}

To design the cyclotron trap assisted positron moderator we performed a detailed simulation with Geant~4 \cite{Agostinelli2003}. The CT magnetic field maps were created with COMSOL and Matlab. Geant~4 was validated to reproduce the correct positron stopping profiles \cite{Dryzek2008} but does not include positron diffusion and the physics of the moderation process.
Therefore, the simulation does not predict the moderation efficiency. To estimate it, we count the fraction of positrons stopping near the surface, in the so-called ejection layer ($\approx$ 100 nm). In fact, those are the ones that have a probability to diffuse to the surface and be emitted as moderated  positrons due to their negative work function in tungsten \cite{MillsReview1988}. 

The simulation suggests that the extraction efficiency by applying +100 V on the source/moderator, with the grid at ground potential, is close to the 94 \% transmission of the grid. The number of fast positrons contributing to the background is expected to be below $10^{-4}$. The simulation of our proof of principle setup for which 2 Ti foils of \SI{1}{\um} thickness where used in order to increase the available positron activity, predicts moderation efficiency of the order of 1 \%.

\begin{figure}[htb]
        \centering
        \includegraphics[width=.5\textwidth]{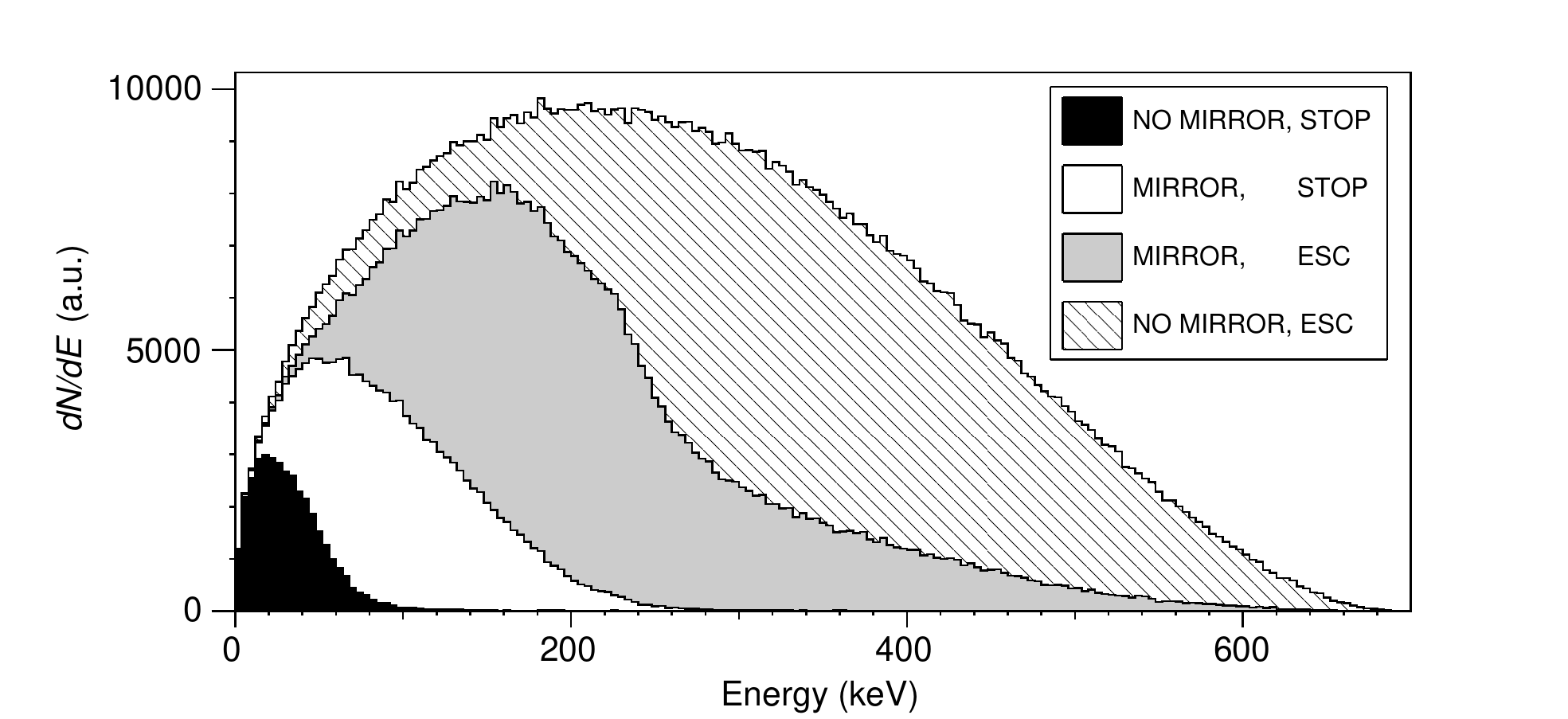}
        \caption{Simulated positron energy spectrum of $^{48}$V separated into fraction of positrons which were not mirrored and stopped inside the moderator or source foil (black area), mirrored at least once and stopped (white area), mirrored at least once and escaped the trap (grey area) and not mirrored and escaped (hatched area).}
        \label{fig:passEtot_creationE}
\end{figure}

\begin{figure}[htb]
        \centering
        \includegraphics[width=.5\textwidth]{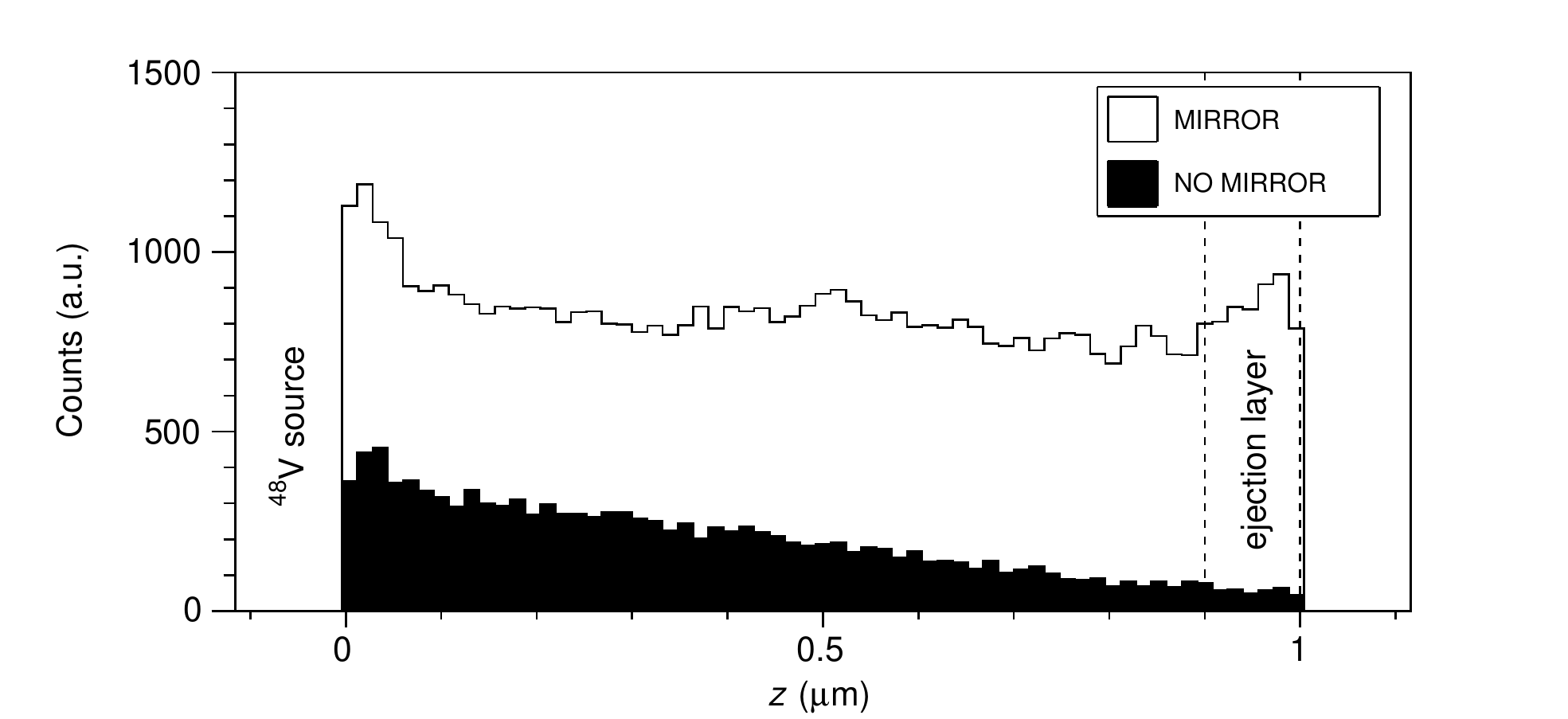}
        \caption{Simulated axial stopping position $z$ split into mirrored (white area) and not mirrored (black area) events. The W moderator foil is ranging from 0 to \SI{1}{\um}. The $^{48}$V source sits on the -$z$ side of the moderator. The ejection layer ranging from 0.9 to \SI{1}{\um} is marked by the dashed lines.}
        \label{fig:decayZ_decayE}
\end{figure}

Figures \ref{fig:passEtot_creationE} and \ref{fig:decayZ_decayE} illustrate the basic principle of assisted positron moderation with a CT. The overall advantage of the trap can be seen in Fig. \ref{fig:passEtot_creationE}. Without a CT just the positrons in the low energy tail of the beta spectrum will stop inside the foil (black area in Fig. \ref{fig:passEtot_creationE}). Of those, only the small fraction stopping in the ejection layer will actually get moderated. With the CT, trapped higher energetic positrons that did not stop in the first pass through the foil will have a chance to stop in one of the subsequent passages thus a part of the beta spectrum that is normally lost can be recycled (white area in Fig. \ref{fig:passEtot_creationE}). The overall enhancement for stopping positrons is a factor of 5. The fact that the moderation efficiency actually improves by a much larger factor is illustrated in Fig. \ref{fig:decayZ_decayE}. The contribution of mirrored positrons (white area), in the ejection layer on the moderator surface opposite to the source, is enhanced compared to the case where no CT is used (black area).  This mimics the so-called reflection ``geometry" which is known to have a higher slow positron yield \cite{Williams2015}.

 The magnetic field of the CT used in this experiment is not strong enough to radially confine all the high energy positrons. The total amount of mirrored positrons are around 66\% for a magnetic field ratio of about 5. As can be seen in Fig. \ref{fig:passEtot_creationE}, only 30 \% of the them finally stop (white area) in the foils and the others escape (grey area). This is due to forward scattering at the foils.
Simulations suggest that with a higher magnetic field ratio and a single source foil, even higher efficiencies are achievable with cyclotron trap assisted positron moderation.


\section{Results}

To measure the background due to the unmoderated positrons and photons reaching the detector, the source and the moderator are kept at ground potential while the extraction grid is biased to +100 V in order to block the moderated positrons. In this configuration, 1 count/s was detected in the electron multiplier.
In the extraction mode, +100 V are applied on the moderator while the grid is at ground, the number of counts in this case was 40 s$^{-1}$. The difference of counts between these two configurations, after correcting for the 70(3) \% detection efficiency of the electron multiplier, is the number of moderated positrons.   
At the time of the experiment the positron activity was $3.2(1)$ kBq, measured with a germanium detector and a calibration source. The division of the counts in the extraction mode by the total positron activity gives a moderation efficiency of the order of $\epsilon \approx 1.8(2)$  \% in fair agreement with what has been estimated from the simulation results.

\section{Conclusions}
The scheme presented here improves the positron moderation of state of the art tungsten moderators by almost two orders of magnitude. It can therefore produce the same output intensity from a much weaker radioactive source. It also yields a factor of two improvement compared to rare gas moderators but is operationally considerably simpler.
 The simulation validated with the measurement predicts that with this technique, using an optimised setup, higher efficiencies can be achieved. Therefore, this technique combined with a few MBq $\beta^+$ source on a \SI{1}{\um} foil would result in a positron flux of $10^4$ positrons/s which is the typical value achieved with standard tungsten moderator based beams.  This opens the possibility to envisage a widespread use of positron beams which are currently available only in few specialized laboratories around the world.
Furthermore, this scheme could be used to increase the moderation efficiency and the yield at high intensity positron facilities thus allowing for further advances in the positron and positronium field.
%

\begin{acknowledgments}
This work has been supported by the Swiss National
Science Foundation under the grant number 200020\_156756.
We would like to thank the PSI and ETH Zurich IPP workshops for their support.
\end{acknowledgments}

\end{document}